\documentclass{cernrep}
\usepackage{graphicx,here}

\begin{document}

\title{BAYESIAN ANALYSIS}

\author{Harrison B. Prosper}

\institute{Department of Physics, Florida  State University, Tallahassee,
Florida 32306, USA}

\def\etal{{\sl et al.}}                 
\def\vs{{\sl vs.}}                      

\def\Journal#1#2#3#4{{#1} {\bf #2} (#4) #3}
\def\NCA{Nuovo Cimento}
\def\NIM{Nucl. Instrum. Methods}
\def\NIMA{{Nucl. Instrum. Methods} A}
\def\NPB{{Nucl. Phys.} B}
\def\PLB{{Phys. Lett.}  B}
\def\PRL{Phys. Rev. Lett.}
\def\APP{Astro. Part. Phys.}
\def\PRD{{Phys. Rev.} D}
\def\PRC{{Phys. Rev.} C}
\def\ZPC{{Z. Phys.} C}
\def\REM{Rev. Mod. Phys.}
\newcommand{\seqn}{\begin{equation}}
\newcommand{\eeqn}{\end{equation}}
\newcommand{\seqna}{\begin{eqnarray}}
\newcommand{\eeqna}{\end{eqnarray}}
\newcommand{\Eq}[1]{Eq.\ (\ref{eq:#1})}
\newcommand{\Eqs}[2]{Eqs.\ (\ref{eq:#1}) and (\ref{eq:#2})}
\newcommand{\lr}[1]{\left ( #1 \right )}
\newcommand{\lrb}[1]{\left [ #1 \right ]}
\newcommand{\bold}[1]{\mbox{\bf #1}}
\newcommand{\clist}[2]{(#1_{1},\ldots,#1_{#2})}
\newcommand{\comb}[2]{\pmatrix{#1\cr#2\cr}}
\newcommand{\R}{{\cal R}}
\newcommand{\N}{{\cal N}}
\newcommand{\C}{{\cal C}}
\newcommand{\Cl}{\mbox{$^{37}Cl$}}
\newcommand{\B}{\mbox{$^{8}B$}}
\newcommand{\Be}{\mbox{$^{7}Be$}}

\newcommand{\prior}[1]{{\rm Prior}(#1|I)}
\newcommand{\like}[2]{{\rm Pr}(#1|#2,I)}
\newcommand{\post}[2]{{\rm Post}(#1|#2,I)}
\newcommand{\prob}[3]{{\rm #1}(#2|#3)}
\newcommand{\Enu}{E_{\nu}}
\newcommand{\ps}{p(\nu|\Enu)}
\newcommand{\psa}{p(\nu|\Enu,a)}

\date{\today}

\maketitle

\begin{abstract}
After making some general remarks, I consider two examples that
illustrate the use of Bayesian Probability Theory. The first is a
simple one, the physicist's favorite ``toy," that provides a forum
for a discussion of the key conceptual issue of Bayesian analysis:
the assignment of prior probabilities. The other example
illustrates the use of Bayesian ideas in the real world of
experimental physics.
\end{abstract}

\section{INTRODUCTION}
\begin{quote}
``We don't know all about the world to start with; our knowledge
by experience consists simply of a rather scattered lot of
sensations, and we cannot get any further without some {\em a
priori} postulates. My problem is to get these stated as clearly
as possible." \vspace{0.2cm}

Sir Harold Jeffreys, in a letter to Sir Ronald Fisher dated 1
March, 1934.
\end{quote}

Scientific inference has led to the surest knowledge we have yet,
paradoxically, there is still disagreement about how to perform
it. The disagreement is both within as well as between camps, the
principal ones being frequentist and Bayesian. If pressed, the
majority of physicists would claim to belong to the frequentist
camp. In practice, we belong to both camps: we are frequentists
when we wish to appear ``objective," but Bayesian when to be
otherwise is either too hard, or makes no sense.
 Until fairly recently, relatively few of us
have been party to the frequentist Bayesian debate.
And society is
all the better for it!
It is
our pragmatism that has cut through the Gordian Knot and allowed
scientific progress.
However, we find ourselves performing
ever more complex inferences that, in some cases, have real world
consequences and we can no longer regard the debate as mere
philosophical musings.
Indeed, this workshop is a testimony to this loss of innocence.

All parties appear, at least, to agree on one thing: probability
theory is a reasonable basis for a theory of inference. But notice
the use of the word ``reasonable." That word highlights the chief
cause of the disagreement: any theory of inference is inevitably
{\em subjective} in the following sense: what one person regards
as reasonable may be considered unreasonable by another and,
unlike scientific theories, we cannot appeal to Nature to decide
which of the many inference theories is best, nor which criteria
are to be used.
 I used to think that
biased estimates were bad. But while some of us strive mightily to
create them others look on bewildered, wondering why on earth we
work so hard to achieve a characteristic they consider irrelevant.

Physicists, quite properly, are deeply concerned about delivering
to the world objective results. Therefore, anything that openly
declares itself to be subjective is viewed with suspicion. Since
Neyman's theory of inference is billed as objective many of us
regard it as reasonable and the Bayesian theory as unfit for
scientific use.  However, when one scrutinizes the Neyman theory,
its ``objectivity" proves to be of a very peculiar sort, as I hope
to show. I then discuss the difficult issue of prior probabilities
by way of a simple model. In the last section, I describe a
realistic Bayesian analysis to illustrate a point: Bayesian
methods are not only fit for scientific use, they are precisely
what is needed to make maximal use of data.

 But first here are some remarks about
probability.

\subsection{What is Probability?}
Probability theory is a mathematical theory about abstractions
called {\em probabilities}. Therefore, to put this theory to work
we are obliged to {\em interpret} these abstractions. At least three
interpretations have been suggested:
\begin{itemize}
\item propensity (Popper)
\item degree of belief (Bayes, Laplace, Gauss, Jeffreys, de Finetti)
\item relative frequency (Venn, Fisher, Neyman, von Mises).
\end{itemize}
In parentheses I have given the names of a few of the proponents.
According to Karl Popper, an unbiased coin, when tossed, has a
propensity of 1/2 to land heads or tails. The 1/2 is claimed to be
a property of the coin. According to Laplace probability is a
measure of the degree of belief in a proposition: given that you
believe the coin to be unbiased your degree of belief in the
proposition ``the coin will land heads" is 1/2. Finally, according
to Venn if the coin is unbiased the relative frequency with which
heads appears in an infinite sequence of coin tosses is 1/2. Venn
seems to have the edge on the other two interpretations since it
is a matter of experience that a coin tossed repeatedly lands
heads about 1/2 the time as the number of tosses, that is, trials,
increases. Every physicist who performs repeated controlled
experiments, either real ones or virtual ones on a computer,
provides overwhelming evidence in support of Venn's
interpretation.

So, which is it to be: degree of belief or relative frequency? The
answer, I believe, is both, which prompts another question: is one
interpretation more fundamental than the other and if so which?
The answer is yes, degree of belief. It is yes for two very
important reasons: one is practical the other foundational. The
practical reason is that we use probability in a much broader
context than that to which the relative frequency interpretation
pertains. It has been amply demonstrated that we perform
inferential reasoning according to rules that are isomorphic to
those of probability theory. Any theory of inference that
dismisses the ``degree of belief" interpretation would be expected
to suffer a severely restricted domain of applicability relative
to the large domain in which probability is used in everyday life.

 The
second reason is that the Venn limit---the convergence of the ratio
of the number of successes to the number of trails---cannot be proved without
appealing to the notion of degree of belief\cite{Jeffreys}.
The issue here is one
of epistemology. Empirical evidence,
even when
overwhelming, does not prove that a thing is true; only that it is
very likely, which is just another way of saying it is very probable. It is
easy to see why a mathematical proof, as commonly understood, cannot be
established. Consider a sequence of trials to test the Standard
Model. Suppose each trial to be a proton
anti-proton collision at the Tevatron. Each trial ends in success
(a top quark is created) or failure. Let $T$ be the number of
trials  and $S$ the number of
successes. Given the top quark mass, the
Standard Model predicts the probability $p$ of successes.
The Standard Model, we note, is a quantum theory. Therefore,
the sequence of successes is strictly non-deterministic,
in a sense in
which a coin toss and a pseudo-random number generator
are not.

However, a necessary (but of course not sufficient) basis for a
mathematical proof of convergence of a sequence to a limit is the
existence of a rule that connects term $T+1$ {\em
deterministically} to $T$. But for quantum theory it is believed
that no such rule exists. What can be and has been proved, by
several people starting with James Bernoulli, is this:
\begin{quote}
If the order of trials is unimportant (that is, the sequence of
trials is {\em exchangeable}), and if the {\em probability} of success
at each trial is the
same, then $S/T \rightarrow p$, as $T
\rightarrow \infty$ with {\em probability} one.
\end{quote}
At this point, I can adopt two attitudes regarding this theorem:
one is that clarity of thought is a virtue; the second is that
clarity of thought is nice but less important than pragmatism. As
a pragmatist I would say that this theorem proves that the Venn
limit exists. But in this case I prefer clarity. Let us,
therefore, be clear about what this theorem actually proves and
what it does not. Bernoulli's theorem does not prove that $S/T$
converges to $p$. Rather it is a statement about 1) the {\em
probability} that $S/T$ converges to $p$ as 2) the number of
trials increases without limit, provided that 3) the order of
trials does not matter and that 4) the {\em probability} at each
trial is the same. Lurking behind these four seemingly innocuous
statements are deep issues that are far beyond the scope of what I
wish to say in this paper. Let me just note that the word
``probability" occurs twice in the statement of Bernoulli's
theorem. If we insist that all probabilities are relative
frequencies then we would have to interpret ``probability of
success at each trial" and ``probability one" as the ``limit with
probability one" of other exchangeable sequences in order to be
consistent. This leads into the abyss of an infinitely recursive
definition. Doubtless, von Mises was well aware of this
difficulty, which may be why he took the existence of the Venn
``limit" as an axiom. However, even if one is prepared to accept
this axiom, I do not think it circumvents the epistemological
difficulty of defining a thing, probability, by making use of the
thing {\em twice} in its definition. As de Finetti\cite{deFinetti}
puts it
\begin{quote}
``In order for the results concerning frequencies to make sense,
it is necessary that the concept of probability, and the concepts
deriving from it which appear in the statements and proofs of
these results, should have been defined and given meaning
beforehand. In particular, a result which depends on certain
events being uncorrelated, or having equal probabilities, does not
make sense unless one has defined in advance what one means by the
probabilities of the individual events."
\end{quote}
I agree.

The alternative interpretation of probability is {\em degree of
belief}. Thus the probability $p$ is our
assessment of the probability of success at each trial, based on
our current state of knowledge. That state of knowledge
could be informed, for example, by the predictions of the Standard
Model. Bernoulli's theorem says that
if our assessment of the probability of success at each trial is
correct, and if our assessment does not change,
then it is reasonable to expect $S/T \rightarrow p$ as
$T \rightarrow \infty$.

But what if our assessment, initially, is incorrect? This poses no
difficulty. As our state of knowledge changes, by virtue of data acquired,
our assessment of the probability of success changes accordingly.
Bayes' theorem shows
how the degree of belief of a coherent reasoner will be updated
to the point
where it closely matches the relative frequency $S/T$.

\subsection{Neyman's Theory}
Neyman rejected the Bayesian use of Bayes' theorem arguing that
the prior probability for a parameter ``has no meaning" when the
latter is an unknown constant. He further argued that even if the
parameters to be estimated could be considered as random variables
we usually do not know the prior probability. With the benefit of
hindsight, we can see that these arguments betray a confusion
about of the notion of degree of belief. Jeffreys\cite{Jeffreys}
frequently lamented the failure of his contemporaries to really
understand what he was talking about. I would note that even
amongst this illustrious gathering the confusion persists. So let
me belabor a point: when one assigns a probability to a parameter
it is not because one deems it sensible to think of the parameter
as if it were a random variable---this is clearly nonsense if the
parameter is in fact a constant. The probability assignments
merely encode one's knowledge (or that of an idealized reasoner)
of the possible values of the parameter.

In his classic paper of 1937\cite{Neyman}, Neyman introduced his
theory of confidence intervals, which he believed provided an
important element of an objective theory of inference. He not only
specified the property that confidence intervals had to satisfy
but he also gave a particular rule for constructing them, although
he left considerable freedom that can be creatively
exploited\cite{Feldman}. Neyman's theory is elegant and powerful.
Nonetheless, his theory is open to criticism. But in order to
raise objections we need to understand what Neyman said.

Imagine an ensemble of trials, or experiments, $\{ E \}$ to each
of which we associate an interval
$[\underline{\theta}(E),\overline{\theta}(E)]$. The ensemble of
experiments yields an ensemble of intervals. Neyman required the
ensemble of confidence intervals to satisfy the following
condition:
\begin{quote}
For every possible {\em fixed} point $(\theta, \alpha)$ in the
parameter space of the problem, where $\theta$ is the parameter of interest
and $\alpha$ denotes all other parameters of the problem
\begin{equation}
{\rm Prob}\{ \theta \in [\underline{\theta}(E),\overline{\theta}(E)]
\} \geq \beta.
\end{equation}
\end{quote}
According to Neyman this probability is to be interpreted as a
relative frequency. Thus,
any set of intervals is an ensemble of {\em
confidence intervals} if the relative frequency with which
the intervals contain the
point $\theta$ is greater than or equal to $\beta$,
for every possible {\em fixed} point in the parameter space regardless of
its dimensionality.
 Neyman's idea is intuitively clear: an interval picked at
{\em random} from such an ensemble, the proverbial urn of
sampling theory, will have a $100\beta$\% chance
of containing the fixed point $\theta$, whatever the
value of $\theta$ and $\alpha$.
This is a remarkable requirement. Here is an example.

Suppose we wish to measure a cross section. Our inference problem
depends upon the following parameters: the cross section $\sigma$,
the efficiency $\epsilon$, the background $b$ and the integrated
luminosity $L$. Consider a {\em fixed} point $(\sigma, \epsilon,
b, L)$ in the parameter space. To this point we associate an
ensemble of confidence intervals, induced by an ensemble of
possible experimental results. Some of these intervals
$[\underline{\sigma}(E),\overline{\sigma}(E)]$ will contain
$\sigma$, others will not. The fraction of intervals, in the
ensemble, that contain $\sigma$ is called the {\em coverage
probability} of the ensemble of intervals. A coverage probability
is associated with every point $(\sigma, \epsilon, b, L)$ of the
parameter space. Moreover, the value of the coverage probability
may vary from point to point. Neyman's key idea is that the
ensembles of intervals should be constructed so that, over the
allowed parameter space, the coverage probability never falls
below some number $\beta$, called the confidence level. Both the
coverage probability and the confidence level are to be
interpreted as relative frequencies.

 The parameter space and its set of ensembles form what
mathematicians call a {\em fibre bundle}. The parameter space is
the base space to each point of which is attached a fibre, that
is, another space, here the ensemble of intervals associated with
that parameter point. Each fibre has a coverage probability, and
none falls below the confidence level $\beta$. Since the fibres
may vary in a non-trivial way from point to point it is not
possible, in general, to construct the fibre bundle as a simple
Cartesian product of the parameter space and a single ensemble of
intervals. In general, a non-trivial fibre bundle is the natural
mathematical description of Neyman's construction. Well natural
if, like me, you like to think of things geometrically!

There are two difficulties with Neyman's idea. The first is
technical. For one-dimensional problems, or for problems in which
we wish to set bounds on {\em all} parameters simultaneously, the
construction of confidence intervals is straightforward. But when
the parameter space is multi-dimensional and our interest is to
set limits on a single parameter no general algorithm is known for
constructing intervals. That is, no general algorithm is known for
eliminating nuisance parameters. In our example, we care only
about the cross-section; we have no interest in setting bounds on
the integrated luminosity. What we do, in practice, is to replace
the nuisance parameters with their maximum likelihood estimates.
The justification for this procedure is the following theorem:
\begin{equation}
-2\log
\frac{Pr(x|\theta,\hat{\alpha})}{Pr(x|\hat{\theta},\hat{\alpha})}
\rightarrow \chi^2, \label{eq:loglike}
\end{equation}
\begin{quote}
as the data sample $x$ grows without
limit, and provided that the maximum likelihood estimates
$\hat{\theta}$
and
$\hat{\alpha}$ lie within the parameter space minus its boundary.
\end{quote}
If our data sample is sufficiently large its likelihood becomes
effectively a (non-truncated) multi-variate Gaussian, and
consequently the distribution of the log-likelihood ratio is
$\chi^2$. Since that distribution is independent of the true
values of the parameters a probability statement about the
log-likelihood ratio can be re-stated as one about the parameter
$\theta$. But, and this is the crucial point, the theorem is
silent about what to do for small samples. Unfortunately, we high
energy physicists insist on looking for new things, so our data
samples are often small. So what are we, in fact, to do? We must
after all publish. Today, with our surfeit of computer time, we
can contemplate a brute-force approach: start with an approximate
set of intervals, computed using \Eq{loglike}, and adjust them
iteratively until they make Neyman happy. But because of the
second difficulty I now discuss the effort seems hardly worth the
trouble.

The second difficulty is conceptual.
 It has been argued at this workshop, and elsewhere\cite{Cousins},
 that the set of published 95\%
 intervals constitute a bona fide ensemble of approximately 95\%
 confidence intervals. Here is the argument. Each published interval
 is
drawn from an urn (that is, an ensemble of experiments if you
prefer a more cheerful allusion) whose confidence level is 95\%.
The fact that each urn is completely different is irrelevant
provided that the sampling probability from each is the same,
namely 95\%. Thus 95\% of the set of published intervals will be
found to yield true statements. And herein lies the beauty of
coverage! The flaw in this argument is this: each published
interval is drawn from an urn that does not objectively exist,
because the ensemble into which an actual experiment is embedded
is a purely conceptual construct not open to empirical scrutiny.
Fisher\cite{Fisher}, not known for fawning over Bayesians, made a
similar point a long time ago:
\begin{quote}
``.. if we possess a unique sample on which significance tests are
to be performed, there is always ... a multiplicity of populations
to each of which we can legitimately regard our sample as
belonging; so the phrase `repeated sampling' from the same
population does not enable us to determine which population is to
be used to define the probability level, for no one of them has
objective reality, all being products of the statistician's
imagination.''
\end{quote}
This is true of our ensemble of experiments.
 Consequently, a few
troublesome physicists, bent on giving the Particle Data Group a
hard time, need merely imagine a different set of urns from which
the published results could legitimately have been drawn and
thereby alter the confidence level of each result!

Of course, the published intervals do have a coverage probability.
My claim is that its value is a matter to be decided by actual
inspection---provided, of course, we know the right answers! It is
not one that can be deduced {\em a priori} for the reason just
given. The fact that I am able to construct ensembles of
confidence intervals on my computer, by whatever procedure, and
verify that they satisfy Neyman's criterion is certainly
satisfying, but in no way does it prove anything empirically
verifiable about the interval I publish. Forgive me for flogging a
sincerely dead horse, but let me state this another way: Since I
do not repeat my experiment, any statement to the effect that the
virtual ensemble simulated on my computer mimics the potential
ensemble to which my published interval belongs is tantamount to
my claiming that if I were to repeat my experiment, then I would
do so such that the virtual and real ensembles matched. Maybe, or
maybe not!

To summarize: A frequentist confidence level is a property of an
ensemble, therefore, its objectivity, or lack thereof, is on par
with the ensemble that defines it.

This whole discussion may strike you as a tad surreal, but I think
it goes to the heart of the matter: many physicists, for sensible
reasons, reject the Bayesian theory and embrace coverage because
it is widely viewed as objective. But as argued above confidence
levels may or may not be objective depending on the circumstances.
Therefore, when confronted with a difficult inference problem our
choice is not between an ``objective" and ``subjective" theory of
inference, but rather between two different subjective theories.
It may be reasonable to continue to insist upon coverage, but not
because it is objective.

After this somewhat philosophical detour it is time to turn to the
real world. But en route to the real world, lest Bayesians begin
to feel uncontrollably smug, I'd like to discuss an instructive
``toy" model that highlights the fact that for a Bayesian life is
hardly a bed of roses\cite{Wasserman}.

\section{THE PHYSICIST'S FAVOURITE TOY}
The typical high energy physics experiment consists of doing a
large number $T$ of similar things---for example, proton
antiproton collisions, and searching for $n$ interesting
outcomes---for example, $t\bar{t}$ production. We invariably
assume that the order of the collisions is irrelevant and that
each interesting outcome occurs with equal probability. Then we
may avail ourselves of the well-known fact that the probability
assigned to $n$ outcomes out of $T$ trials, with our assumptions,
is binomial. Since $n << T$, this probability can be approximated
by a Poisson distribution
\begin{equation}
\like{n}{\mu} = \frac{e^{-\mu} \mu^{n}}{n!}, \label{eq:poisson}
\end{equation}
and thus do we arrive at the physicist's favourite toy. The symbol
$I$ denotes all prior information and assumptions that led us to
this probability assignment. Here, it is introduced for
pedagogical reasons; to remind us of the fact that {\em all}
probabilities are conditional. We shall assume that our aim is to
infer something about the Poisson parameter $\mu$, given that we
have observed $n$ events. Just for fun, we'll give this problem to
each workshop member. Naturally, being physicists, each of us
insists on parameterizing this problem as we see fit, but in the
end when we compare notes we shall do so in terms of the parameter
$\mu$,  by transforming to that parameter.

There are, of course, infinitely many ways to parameterize a
likelihood function and the Poisson likelihood is no exception.
For simplicity, however, let's assume that each of us uses a
parameter $\mu_p$ related to $\mu$ as follows
\begin{equation}
\mu_p = \mu^p.
\label{eq:mup}
\end{equation}
``$p$" for physicist if you like! In terms of the parameter
$\mu_p$ \Eq{poisson} becomes
\begin{equation}
\like{n}{\mu_p} = \frac{e^{-\mu_p^{1/p}} \mu_p^{n/p}}{n!},
\label{eq:poissonp}
\end{equation}
which, we note, does not alter the probability assigned to $n$.

From Bayes' theorem
\begin{equation}
\post{\mu_p}{n} = \frac{\like{n}{\mu_p}
\prior{\mu_p}}{\int_{\mu_p} \like{n}{\mu_p} \prior{\mu_p}},
 \label{eq:defpostp}
\end{equation}
each of us can make inferences about our parameter $\mu_p$, and
hence $\mu$. Of course, no one can proceed without specifying a
prior probability $\prior{\mu_p}$. Unfortunately, being mere
physicists we do not know what its form should be. But since we
are all in the same state of knowledge regarding our parameter,
coherence would seem to demand that we use the same functional
form. So without a shred of motivation let's try the following
form for the prior probability
\begin{equation}
\prior{\mu_p} = \mu_p^{-q} d\mu_p. \label{eq:aprior}
\end{equation}
Although this prior is plucked out of thin air, it is actually
more general than it appears because, in principle, $q$ could be
an arbitrarily complicated function of $p$. Now each of us is in a
position to calculate, assuming that the allowed parameter space
for $\mu_p$ is $[0,\infty)$. We each find that
\begin{equation}
\post{\mu_p}{n} = \frac{e^{-\mu_p^{1/p}} \mu_p^{n/p-q}
d\mu_p}{p\Gamma(n-pq+p)}.
 \label{eq:postp}
\end{equation}
But as agreed, each of us transforms our posterior probability to
the parameter $\mu$ using \Eq{mup}. Thus we obtain, from
\Eq{postp},
\begin{equation}
\post{\mu}{n} = \frac{e^{-\mu} \mu^{n-pq+p-1}
d\mu}{\Gamma(n-pq+p)}.
 \label{eq:postmu}
\end{equation}
Unfortunately, something is seriously amiss with the family of
posterior probabilities represented by \Eq{postmu}: each of us has
ended up  making a different inference about the same parameter
$\mu$! We can see this more clearly by computing the $r$th moment
\begin{eqnarray}
\label{eq:mr}
 m_r & \equiv & \int_{\mu} \mu^r \post{\mu}{n} \\ \nonumber
    & = & \Gamma(n-pq+p+r)/\Gamma(n-pq+p),
\end{eqnarray}
of the posterior probability $\post{\mu}{n}$. The moments clearly
depend on $p$, that is, on how we have chosen to parameterize the
problem.

What does a Bayesian have to say about this state of affairs? Is
it a problem? I would say yes, it is. But there are some Bayesians
who call themselves ``subjective Bayesians" and others who believe
themselves to be ``objective Bayesians." I confess that these
terms leave me a bit baffled. The latter term because it seems to
be an oxymoron and the former because it seems to be superfluous.
The fundamental Bayesian pact is this: The prior probability is an
encoding of a state of knowledge; as such it is a subjective
construct. That construct may encode one's personal state of
knowledge or belief, and that's a fine thing to do and is very
powerful. But it may also encode a state of knowledge that is not
specifically yours and that too is just fine. The issue is one of
encoding a state of knowledge: Are there any desiderata that
should be respected? The subjectivist is probably inclined to say
no: simply choose the parameterization that makes sense for you
and associate a prior, declare it to be supreme, and force all
other priors to differ from yours in just the right way to render
an inference about $\mu$ unique. So a ``subjective" Bayesian would
presumably reject \Eq{aprior}.

I believe that to make headway, we should entertain some further
principles. They should not degenerate into dogma but should serve
as a lantern in the dark. Here are two possible principles:
\begin{itemize}
\item Possible Principle 1: For the same likelihood and the same
form of prior we should obtain the same inferences.
\item Possible Principle 2: The moments of the posterior
probability should be finite.
\end{itemize}
Let's apply these tentative principles to the moments in \Eq{mr}.
Principle 1 says that each of us should make the same inferences
about $\mu$, that is, the moments ought not to depend on the whim
of a workshop member; it ought not to depend on $p$. Principle 2
says that $m_r < \infty$. Together these principles imply that
\begin{equation}
-pq + p = a > 0,
\end{equation}
where $a$ is a constant. This leads to the following prior
\begin{equation}
\prior{\mu_p} = \mu_p^{a/p-1} d\mu_p.
\end{equation}
But we didn't quite make it; our principles are insufficient to
uniquely specify a value for the constant $a$. We need something
more. Here is something more, suggested by Vijay
Balasubramanian\cite{Bala}:
\begin{itemize}
\item Possible Principle 3: When in doubt, choose a prior that
gives equal weight to
all likelihoods indexed by the same parameters.
\end{itemize}
That is, impose a {\em uniform} prior on the space of
distributions. This requirement is a much more reasonable one
(here is that word again) than imposing uniformity on the space of
parameters because the space of distributions is invariant,
whereas that of parameters is not. The space of distributions is
akin to a space containing invariant objects like the vectors in a
vector space, whereas the parameter space is analogous to the
non-invariant space of vector coordinates. In our case, we impose
a uniform prior on the space inhabited by Poisson distributions.
Balasubramanian has shown that a uniform prior on the space of
distributions induces, locally, a Riemannian metric whose
invariant measure is determined by the Fisher Information, $F$.
For our toy model the invariant measure is
\begin{equation}
\label{eq:jeffprior} \prior{\mu_p} = F^{1/2} d\mu_p,
\end{equation}
where
\begin{equation}
F(\mu_p) = -\left < \frac{d^2 \log \like{n}{\mu_p}}{d\mu_p^2}
\right >.
\end{equation}
Equation~(\ref{eq:jeffprior}) is called the {\em Jeffreys prior}.
It gives $a = 1/2$ and thus uniquely specifies the form of the
prior probability. Possible Principle 3 is a generalization of
Possible Principle 1. Thus we conclude that the prior probability
that forces us all to make the same inference, regardless of how
we choose to parameterize the problem, is
\begin{equation}
\label{eq:prior} \prior{\mu_p} = \mu_p^{-\frac{1}{2}(2-p)} d\mu_p.
\end{equation}

This is all very tidy. However, when Jeffreys\cite{Jeffreys}
applied his general prior probability to the Gaussian, treating
both its mean and standard deviation together he got a result he
did not like. He therefore suggested another principle:
\begin{itemize}
\item Possible Principle 4: If the parameter space can be
partitioned into subspaces that, {\em a priori}, are considered
independent then the general prior should be applied to each
subspace separately.
\end{itemize}
This gave him a prior he liked. Alas, for a Bayesian life is not
easy. While the frequentist struggles with justifying the use of a
particular non-objective ensemble the Bayesian struggles to
justify why some set of additional principles for encoding minimal
prior knowledge is reasonable. Meanwhile, the ``subjective
Bayesian" says this is all a mere chasing after shadows. And so it
goes!

\section{THE READ WORLD}
The foregoing discussion might suggest to ``abandon all hope yea
who enter" the real world of inference problems. Fortunately, it
is not quite so bleak. The real world imposes some very severe
constraints on what we can reasonably be expected to do. For one
thing, the lifetime of a physicist is finite, indeed, short when
compared with the age of the universe. Technical resources are
also finite. And then there is competition from fellow physicists.
Finally, uncertainty in abundance is the norm. Perhaps with enough
deep thought all inference problems can be solved in a pristine
manner. In practice, we are forced to exercise a modicum of
judgement when undertaking any realistic analysis. We introduce
approximations as needed, we side-step difficult issues by
accepting some conventions and we rely upon our ability not to get
lost amongst the trees. But when I reflect on what must be done to
measure, say, the top quark mass, a problem replete with
uncertainties in the jet energy scale, acceptance, background,
luminosity, Monte Carlo modeling to name but a few, it strikes me
as desirable to have a coherent and intuitive framework to think
about such problems. Bayesian Probability Theory provides
precisely such a framework. Moreover, it is a framework that
mitigates our propensity to get confused about statistics when the
going gets tough. The second example I discuss shows that real
science can be done in spite of prior anxiety\cite{Wasserman}.

\subsection{Measuring the Solar Neutrino Survival Probability}

\begin{figure}[htbp]
\begin{center}
\includegraphics[width=8.5cm,angle=-90]{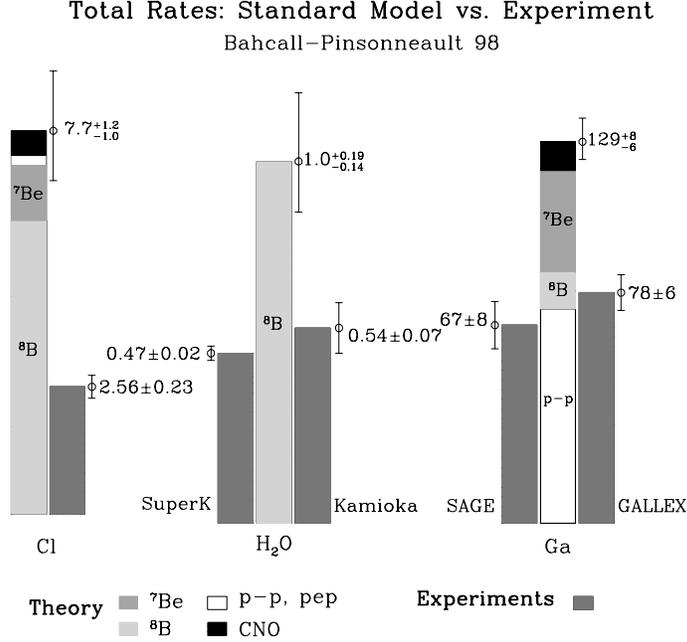}
\caption{Predictions of the 1998 Standard Solar Model of Bahcall
and Pinsonneault relative to data presented at Neutrino 98.
Courtesy J.N.~Bahcall.} \label{fig:rates}
\end{center}
\end{figure}

It has been known for over a quarter of a century that fewer
electron neutrinos are received from the Sun than expected on the
basis of the Standard Solar Model
(SSM)\cite{review,history,bp98,bp95,ssm93}.
 This is the famous solar neutrino
problem. Figure~\ref{fig:rates} summarizes the situation as of
Neutrino 98.
If the SSM is correct---and there is very strong evidence in its
favour\cite{helio}, then the inevitable conclusion is that a
fraction of the electron neutrinos created in the solar core are
lost before they reach detectors on Earth. The loss of electron
neutrinos is parameterized by the {\em neutrino survival
probability}, $\ps$, which is the probability that a solar
neutrino $\nu$ of energy $\Enu$ arrives at the Earth.

Several loss mechanisms have been suggested, such as the
oscillation of electron neutrinos to less readily observed states
such as muon, tau  or sterile neutrinos\cite{gribov,wolfenstein}.
Many $\chi^2$-based analyses have been performed to estimate model
parameters\cite{hata, lui, parke}. To the degree that a fit to the
solar neutrino data is good it provides evidence in favour of the
particular new physics that has been assumed. From this
perspective, solar neutrino physics is yet another way to probe
physics beyond the Standard Model.

 But I'd like to address a more modest question:
 What do the data tell us about the solar neutrino
 survival probability independently of any particular model of new physics?
  We can provide a complete answer by computing
 the posterior probability of different
 hypotheses about the value of the survival probability, for a given neutrino
 energy\cite{bhat,gates}. Our Bayesian analysis is comprised of four components
\begin{itemize}
\item The model
\item The data
\item The likelihood
\item The prior
\end{itemize}
First we sketch the model. (See Ref.~\cite{bhat} for details.)

\begin{figure}[htbp]
\begin{center}
\includegraphics[width=8.5cm,angle=-90]{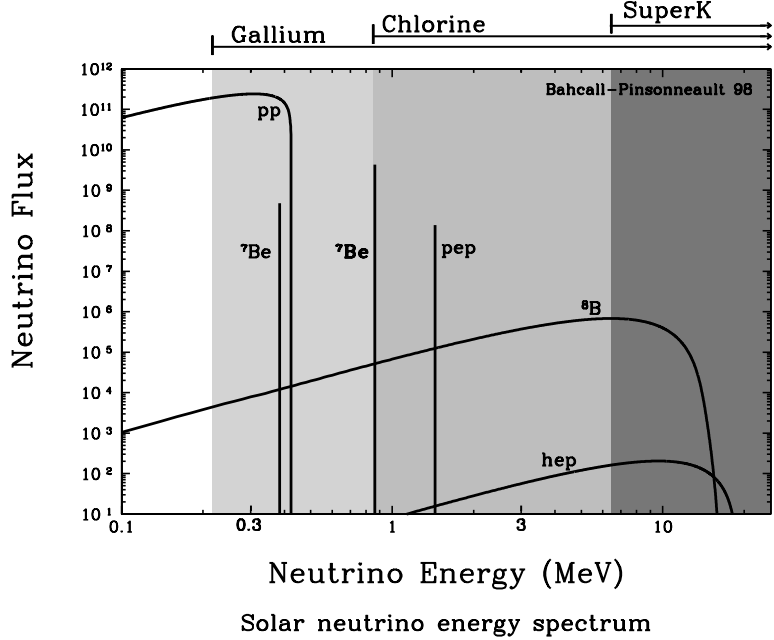}
\caption{Solar neutrino energy spectrum as predicted by the
Bahcall-Pinsonneault 1998 Standard Solar Model, including the
neutrino energy thresholds for different solar neutrino
experiments. Courtesy J.N.~Bahcall.} \label{fig:spectrum}
\end{center}
\end{figure}

The solar neutrino capture rate $S_i$ on chlorine and gallium can
be written as
\begin{equation}
S_i = {\sum_j} \Phi_j \int  \ps \sigma_i(E_{\nu}) \phi_j(E_{\nu})
d\Enu, \label{eq:radio}
\end{equation}
where $\Phi_j$ is the total flux from neutrino source $j$,
$\phi_j$ is the normalized neutrino energy spectrum and $\sigma_i$
is the cross section for experiment $i$. The predicted spectrum,
plus experimental energy thresholds, are shown in
Fig.~\ref{fig:spectrum}. The full spectrum consists of eight
components (of which six are shown in Fig.~\ref{fig:spectrum}),
with total fluxes $\Phi_1$ to $\Phi_8$\cite{bp98}.

\begin{figure}[htbp]
\begin{center}
\includegraphics[width=8.5cm]{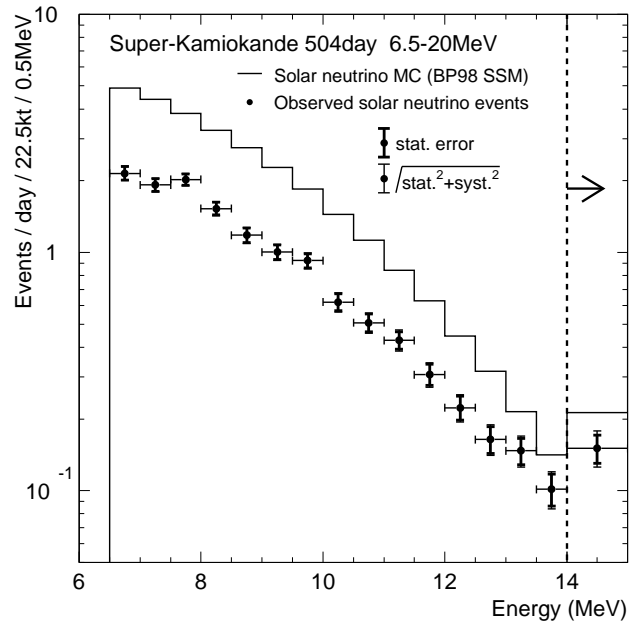}
\caption{Electron recoil spectrum measured by Super-Kamiokande
compared to spectrum predicted by the Bahcall-Pinsonneault 1998
Standard Solar Model. From Ref.~\cite{superk}.} \label{fig:recoil}
\end{center}
\end{figure}

The Super-Kamiokande experiment\cite{neutrino98} measures the
electron recoil spectrum arising from the scattering of the $^8B$
neutrinos (plus higher energy neutrinos) off atomic electrons.  We
shall use the electron recoil spectrum reported at Neutrino 98.
The spectrum spans the range 6.5 to 20~MeV. Light water
experiments, like Super-Kamiokande, are sensitive to all neutrino
flavors but do not distinguish between them. There are, therefore,
two possibilities: the $\nu_e$ deficit could be caused by $\nu_e$
conversions to $\nu_x$, where $x$ is either $\mu$ or $\tau$. If so
the measured neutrino flux would be the sum of these flavors. If,
however, the $\nu_e$ are simply lost without a trace, for example
because of conversion into sterile neutrinos, then the measured
flux would be comprised of $\nu_e$ only. Like the rates for the
radiochemical experiments, the measured electron recoil spectrum
is linear in the neutrino survival probability. The data are shown
in Fig.~\ref{fig:recoil}.

\begin{figure}[htbp]
\begin{center}
\includegraphics[width=8.5cm]{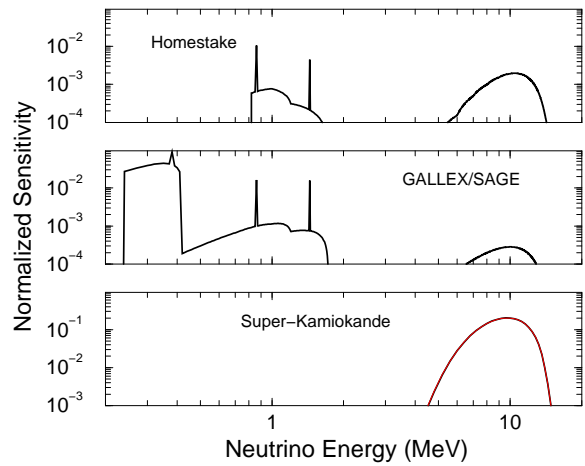}
\caption{Spectral sensitivity as a function of the neutrino
energy. From Ref.~\cite{bhat}.} \label{fig:sensitivity}
\end{center}
\end{figure}

For solar neutrino experiments, a reasonable definition of
sensitivity is the product of the cross section times the
spectrum\cite{bhat}. This quantity is plotted in
Fig.~\ref{fig:sensitivity}. Two points are noteworthy: each
experiment is sensitive to different parts of the neutrino energy
spectrum and there are regions in neutrino energy where the
sensitivity is essentially zero. We should anticipate that these
facts will constrain what we are able to learn about the neutrino
survival probability from the current solar neutrino data.

Since we do not know the cause of the solar neutrino deficit,
let's adopt a purely phenomenological approach to the survival
probability. Guided by the results from previous
analyses~\cite{hata,lui,parke,cbhat} we write the survival
probability as a sum of two finite Fourier series:
\begin{eqnarray}
\label{eq:parametric}
 \psa & = & \sum_{r=0}^{7}
a_{r+1} \mbox{cos}(r\pi E_{\nu}/L_1)  / (1 +
\mbox{exp}[(E_{\nu}-L_1)/b])
\\ \nonumber
& + & \sum_{r=0}^{3} a_{r+9} \mbox{cos}(r \pi E_{\nu}/L_2),
\end{eqnarray}
where now we explicitly note the fact that the survival
probability depends upon the set of parameters $a$.
The first term in \Eq{parametric} is defined in the interval 0.0
to $L_1$ MeV---and suppressed beyond $L_1$ by the exponential. The
second term spans the interval 0.0 to $L_2$ MeV. We have divided
the function this way to model a survival probability that varies
rapidly in the interval 0.0 to $L_1$ and less so elsewhere. The
parameters $L_1$, $L_2$ and $b$ are set to 1.0, 15.0 and 0.1~MeV,
respectively.

We now consider the likelihood function $\like{D}{H}$, where $H$
denotes the hypothesis under consideration. The likelihood is
assumed to be proportional to a multi-variate Gaussian
$g(D|S,\Sigma)$, where $D \equiv (D_1,\ldots,D_{19})$ represents
the 19 data---3 rates from the chlorine and gallium experiments
plus 16 rates from the binned Super Kamiokande electron recoil
spectrum (Fig.~\ref{fig:recoil}); $\Sigma $ denotes the
$19\times19$ error matrix for the experimental data and $S \equiv
(S_1,\ldots,S_{19})$ represents the predicted rates.

The remaining ingredient is the prior probability. First we assess
our state of knowledge. There are two sets of parameters to be
considered: the total fluxes $(\Phi_1,\ldots,\Phi_8)$ and the
survival probability parameters $(a_1,\ldots,a_{12})$. The
hypotheses under consideration concern the values of these two
sets of parameters. The Standard Solar Model provides predictions
$\Phi^0 \equiv (\Phi_1^0,\ldots,\Phi_8^0)$ for the total fluxes,
together with estimates of their {\em theoretical} uncertainties.
So here is an analysis that must deal with theoretical
uncertainties in some sensible way. I do not know how such a thing
can be addressed in a manner consistent with frequentist precepts.
For a Bayesian uncertainty is, well, uncertainty, regardless of
provenance; therefore, every sort can be treated identically. We
represent our state of knowledge regarding the fluxes by a
multi-variate Gaussian prior probability $\prior{\Phi} =$
$g(\Phi|\Phi^0,\Sigma_{\Phi})$, where $\Phi^0$ is the vector of
flux predictions and $\Sigma_{\Phi}$ is the corresponding error
matrix\cite{bp98}.

 Unfortunately, we know very
little about the parameters $a_1, \ldots,a_{12}$, so we shall
short-circuit discussion by taking, as a matter of convention, the
prior probability for $a$ to be uniform. In practice, any other
plausible choice makes very little difference to our conclusions.
We may even find that a uniform prior for $a$ is consistent with
the generalized Jeffreys prior. Thus we arrive at the following
prior for this inference problem:
\begin{eqnarray}
\prior{a,\Phi} & = & {\rm Prior}(a|\Phi,I) \prior{\Phi} \\
\nonumber
    & = & da \prior{\Phi},
\end{eqnarray}
where $I$ now includes the prior information from the Standard
Solar Model.

\begin{figure}[htbp]
\begin{center}
    \begin{minipage}[t]{0.46\linewidth}
    \includegraphics[width=7cm,angle=-90]{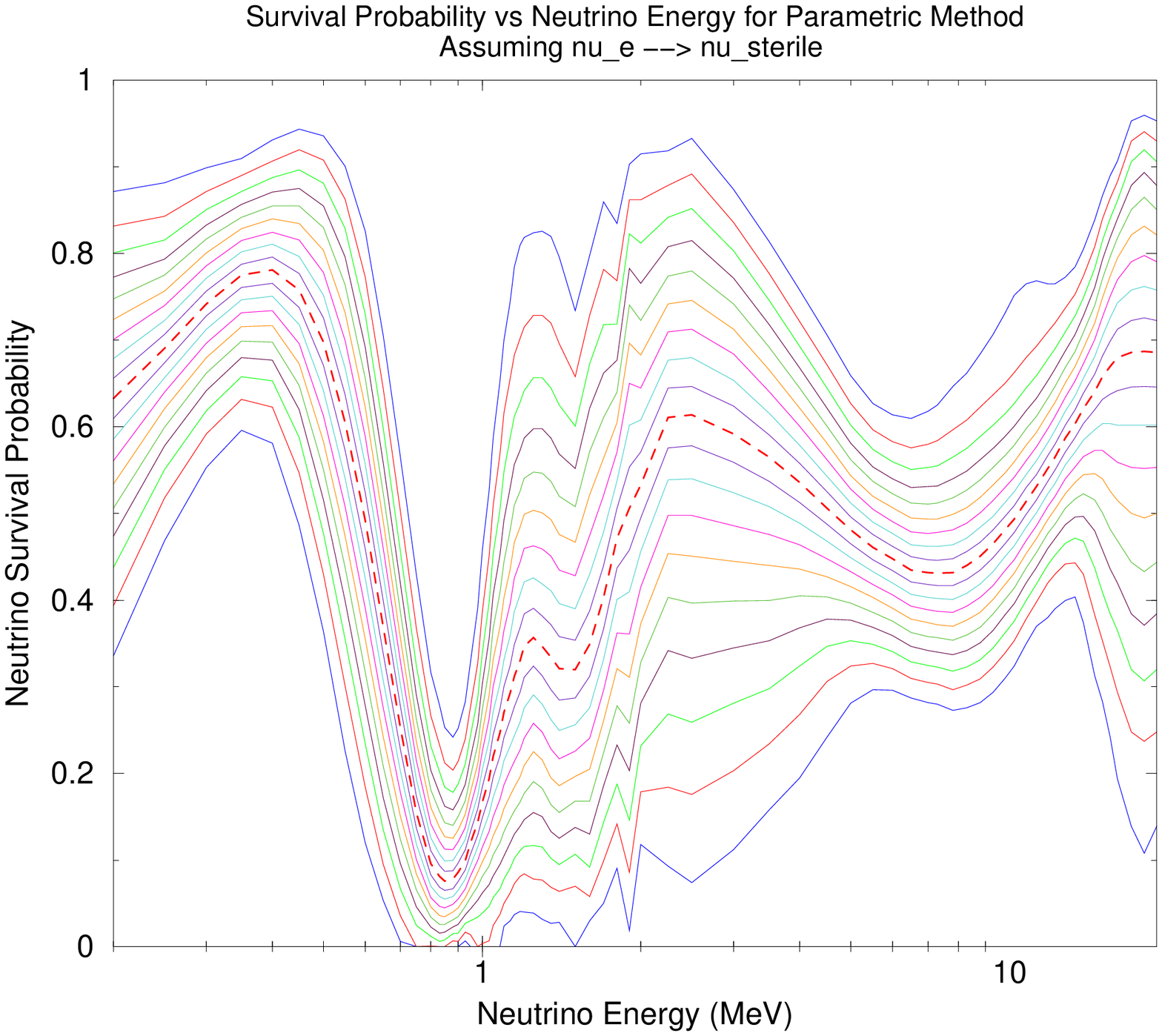}
    \end{minipage}
    \begin{minipage}[t]{0.46\linewidth}
    \includegraphics[width=7cm,angle=-90]{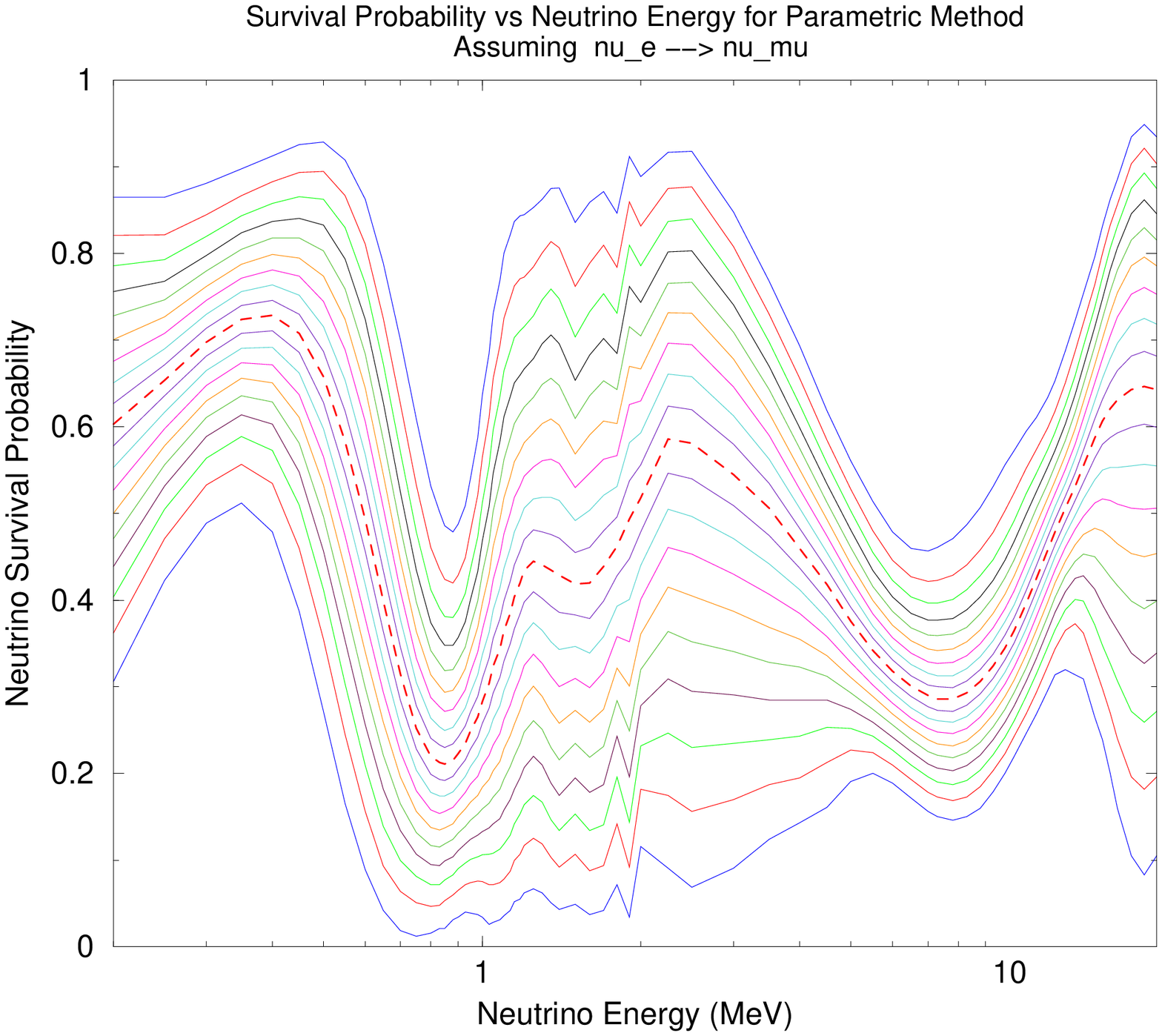}
    \end{minipage}
\end{center}
\caption{ Survival probability {\it vs} neutrino energy assuming
the  neutrino flux consists of $\nu_e $ only (left plot) and
$\nu_e $ to active neutrinos (right plot).} \label{fig:parme}
\end{figure}

Now we can calculate! The posterior probability is given by
\begin{equation}
\post{a,\Phi}{D} = \frac{\like{D}{a,\Phi} \prior{a,\Phi}}
            {\int_{a,\Phi}\like{D}{a,\Phi} \prior{a,\Phi}  }.
\end{equation}
But since we aren't really interested in the total fluxes
probability theory dictates that we just marginalize  (that is,
integrate) them away to arrive at the quantity of interest
$\post{a}{D}$. Actually, what we really want is the probability of
the survival probability for a given neutrino energy $\Enu$! That
is, we want
\begin{equation}
    \post{p}{D} = \int_{a} \delta(p - \psa) P(a|D,I).
\end{equation}
 Figure~\ref{fig:parme} shows contour plots of $\post{p}{D}$ for the two
 cases considered, conversion to sterile and active neutrinos.

Our Bayesian analysis has produced a result that, intuitively,
makes a lot of sense. As expected, given the sensitivity plot in
Fig.~\ref{fig:sensitivity}, our knowledge of the survival
probability is very uncertain between 1 and 5 MeV. In fact, the
survival probability is tightly constrained in only two narrow
regions: in the \Be\ region just below 1 MeV and another at around
8 MeV, near the peak of the \B\ neutrino spectrum. For neutrino
energies above 12 MeV or so, the survival probability is basically
unconstrained by current data.

\section{SUMMARY}
It has been claimed by some at this workshop that Bayesian methods
are of limited use in physics research. This of course is not true
as I hope to have shown. Bayesian methods are, however, explicitly
subjective and this may give one pause. I have argued that
frequentist methods are not nearly as objective as claimed. While
Bayesians cannot avoid the irreducible subjectivism of prior
probabilities, frequentists cannot avoid the use of ensembles that
do not objectively exist. Frequentists struggle with any
uncertainty that does not arise from repeated sampling, like
theoretical errors, while for Bayesians uncertainty in all its
forms is treated identically. On the other hand, some Bayesians
struggle to convince us that a particular choice of prior is
reasonable, while frequentists look on in amusement. The point is
neither approach is free from warts. But, of the two approaches to
inference, I would say that the Bayesian one has more to offer, is
easier to understand, has greater conceptual cohesion and, the
most important point of all, more closely accords with the way we
physicists think\cite{bayes}. And this is real reason why it
should be embraced.

\vskip1cm \noindent

\section*{ACKNOWLEDGEMENTS}

 I wish to thank the organizers for hosting this most enjoyable workshop.
 It was a particular pleasure for me to meet again my dear friend, and
 intellectual sparring partner, Fred James who must take all the
 credit for arousing my interest in this arcane subject.
 I thank my colleagues Chandra Bhat, Pushpa Bhat
and Marc Paterno with whom the solar neutrino work was done, John
Bahcall for providing the latest theoretical information and
Robert Svoboda for providing the 1998 Super-Kamiokande data in
electronic form. This work was supported in part by the U.S.
Department of Energy.

\end{document}